\documentclass[aps,prb,twocolumn,footinbib,amsmath,amssymb,amsfonts,superscriptaddress]{revtex4-1}

\usepackage[varg]{txfonts}
\usepackage{amsmath}
\usepackage{graphicx}
\usepackage{subfigure}
\usepackage[abs]{overpic}
\usepackage{color}

\newcommand{\h}[1]{{#1}^{\dagger}} 
\newcommand{\cc}[1]{{#1}^{*}}
\newcommand{\cb}[1]{\bar{#1}}

\newcommand{\vecc}[1]{\vec{#1}}

\newcommand{\im}{{\rm Im}}

\newcommand{\U}{{\rm U}}

\begin{document}

\title{Hidden and antiferromagnetic order as a rank-5 superspin in URu$_{\textrm{\bf 2}}$Si$_{\textrm{\bf 2}}$ 
}
\author{Jeffrey G. Rau}
\affiliation{Department of Physics, University of Toronto, Toronto, Ontario M5S 1A7, Canada}
\author{Hae-Young Kee}
\email[Electronic Address: ]{hykee@physics.utoronto.ca}
\affiliation{Department of Physics, University of Toronto, Toronto, Ontario M5S 1A7, Canada}
\affiliation{Canadian Institute for Advanced Research/Quantum Materials Program, Toronto, Ontario MSG 1Z8, Canada}

\date{\today}

\begin{abstract}
We propose a candidate for the hidden order
in URu$_2$Si$_2$: a rank-5 $E$ type spin density wave
between Uranium $5f$ crystal field doublets $\Gamma^{(1)}_7$ and $\Gamma^{(2)}_7$,
breaking time reversal and lattice tetragonal symmetry in a manner consistent
with recent torque measurements [R. Okazaki et al, Science {\bf 331}, 439 (2011)]. We argue that 
coupling of this order
parameter to magnetic probes can be hidden by crystal
field effects, while still having significant effects on transport, thermodynamics and 
magnetic susceptibilities. In a simple tight-binding model for the
heavy quasiparticles, we show the connection between the hidden order and 
antiferromagnetic phases arises since
they form different components of this single rank-5 pseudo-spin vector.
Using a phenomenological theory, we show the
experimental pressure-temperature phase diagram 
can be qualitatively reproduced by tuning terms which break pseudo-spin
rotational symmetry. As a test of our proposal, we predict the presence of small magnetic moments in the basal
plane oriented in the $[110]$ direction ordered at the wave-vector  $(0,0,1)$.
\end{abstract}


\maketitle
\section{Introduction}
The nature of the low temperature ordered phase found in the heavy fermion
compound URu$_2$Si$_2$ has defied explanation for over 25 years.
While the transition into this hidden order (HO) phase
appears quite conventional based on the 
effects on thermodynamic\cite{franse-1986,mydosh-1985},
magnetic \cite{mydosh-1985,broholm-1991,ramirez-1992} and
transport phenomena\cite{mydosh-1985,franse-1987,dawson-1989}
, the order parameter itself remains elusive.
 This stands in contrast
to the significant entropy release\cite{mydosh-1985} across the transition,
indicating a strong ordering, inconsistent with the weak antiferromagnetic
moments that perplexed early studies.
Heroic efforts have brought the full complement of experimental techniques
to bear upon this problem, from local probes such as $\mu$SR\cite{amato-2004}, NMR\cite{matsuda-2001,mydosh-2001}
and NQR\cite{saitoh-2005}, surface probes\cite{arpes,softarpes,qpi}, scattering studies using elastic and inelastic neutrons\cite{broholm-1987,broholm-1991,villaume-2008} and 
resonant X-rays\cite{igarashi-2005,floquet-2011} as well as explorations of the material through
pressure, both hydrostatic\cite{amitsuka-1999,hassinger-2008,villaume-2008} and uniaxial\cite{yokoyama-2005} and high magnetic fields\cite{mydosh-2007}. 
Key facets of the problem illuminated by these studies include a
pressure induced antiferromagnetic phase (AF) and a
superconducting phase that arises out of the HO phase.
Similarities between the HO and AF phases,
such as their Fermi surfaces\cite{nakashima-2003,hassigner-2010}
suggest a common underlying mechanism.
While great
progess has been made, the central mystery of the identity of the HO still remains.

Recently, an important clue to the nature of the HO has been
seen in magnetic torque experiments\cite{okazaki-torque}. Through
a measurement of the off-diagonal magnetic susceptibility on
small samples, it was found that the HO phase breaks the rotational
symmetry of the crystal spontaneously in the $[110]$ direction. 
Lack of detectable lattice distortions\cite{lattice-distortion,lattice-distortion-2} 
suggest this symmetry breaking is purely
an electronic phenomenon.
This is inconsistent with many earlier theoretical proposals
for the HO (see \cite{review-article} for a review) and thus has attracted
a great deal of interest. New proposals to explain these 
observations include, among others, spin-nematic states\cite{fujimoto-nematic},
dynamical symmetry breaking\cite{oppeneer-dft,oppeneer-dft-2}, staggered spin-orbit
coupling order\cite{das-2012} and
hastatic order \cite{chandra-unpublished}.

In this article, we propose a candidate for the hidden order
in URu$_2$Si$_2$ as a rank-5 $E$ type 
spin density wave
between $5f$ crystal field doublets $\Gamma^{(1)}_7$ and $\Gamma^{(2)}_7$. This breaks both 
time reversal and the lattice point group symmetry $D_{4h}$ in a manner consistent
with the torque result. We argue that the expected coupling of this order
parameter to magnetic probes, such as neutrons, can be effectively hidden by crystal
field effects in the magnetic moment, while still contributing to second-order correlations such 
as susceptibilities. This would manifest as a small 
moment in the basal plane oriented along the $[110]$ direction ordered
at wave-vector $(0,0,1)$.
In a simple tight-binding model for the itinerant
heavy quasiparticles, we show the close relation between the HO and AF phases arises
due to an approximate degeneracy between the $E$ type ($x$ and $y$ components)
and $A_2$ type ($z$ component) of the rank-5 pseudo-spin vector\footnote{
In free space these order parameters are composed mainly
of rank-5 operators mixed with rank-3 operators.}
\begin{equation}
  \label{order-parameter}
  \vecc{\phi} = \left\langle \frac{1}{N}\sum_i e^{i\vecc{Q}\cdot\vecc{r}_i} \left(
    \frac{\vecc{S}_{12}(i) + \vecc{S}_{21}(i)}{\sqrt{2}}\right) \right\rangle.
\end{equation}
where $\vecc{Q} = 2 \pi/c \hat{z} \equiv (0,0,1)$ and we have defined generalized spin
operators $\vecc{S}_{\alpha\beta} = \frac{1}{2}\h{f}_\alpha \vecc{\sigma} f_\beta$
where $\alpha,\beta=1,2$ indicate the $\Gamma^{(7)}$ doublet.
These doublets are related
to the $5f$ $J_z$ eigenstates as 
  $f_{1\pm} = \cos{\theta}\ f_{\pm \frac{5}{2}} + \sin{\theta}\ f_{\mp \frac{3}{2}}$ and 
  $f_{2\pm} = -\sin{\theta}\ f_{\pm \frac{5}{2}} + \cos{\theta}\ f_{\mp \frac{3}{2}}$
where $\theta$ is determined by the relative crystal field
strengths \cite{thalmeier-landau}. 
 The similarity of
the Fermi surfaces arises as the
breaking of the pseudo-spin symmetry is inoperative in the $k_x-k_y$ plane, leading to similar
cross-sections and thus oscillations. Using a phenomenological theory, we show the
experimental phase diagram as a function of pressure and temperature
can be qualitatively reproduced by tuning either the hopping which
breaks pseudo-spin rotational symmetry
or the effective coupling constants. As a consequence of
this pseudo-spin symmetry breaking, the Goldstone mode connecting
the HO and AF becomes gapped, leading to a resonance a finite frequency
in the magnetic response.
Specific heat and magnetic susceptibilities 
are also presented, and experimental means to test our proposal 
are discussed.
\begin{figure}[t]
\subfigure[\ $k_x - k_y$ plane]{
\begin{overpic}[width=0.4\columnwidth]
{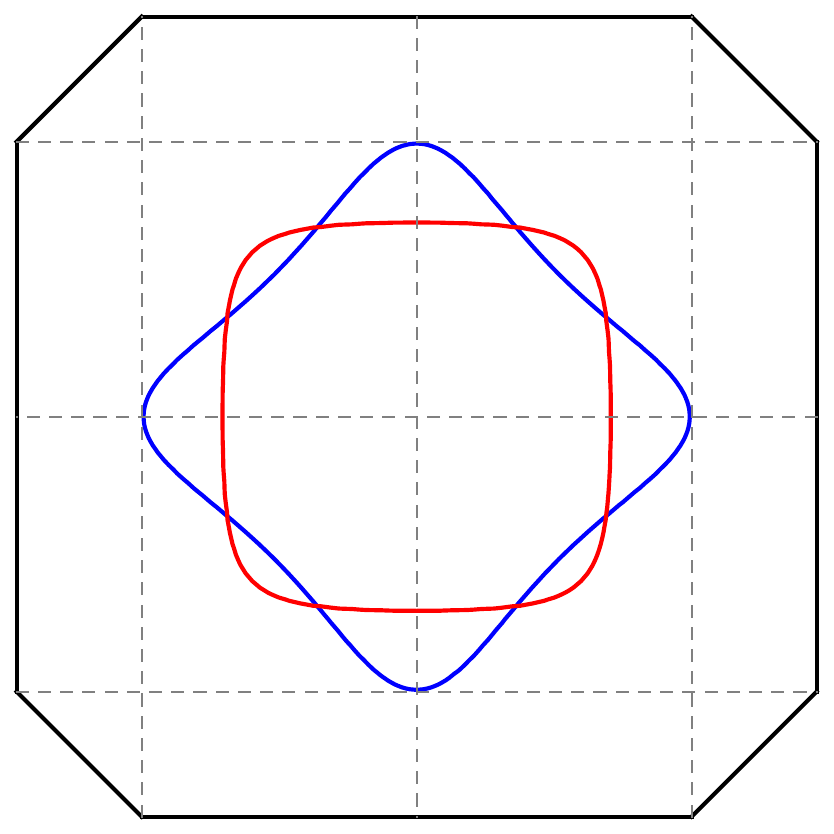}
\put(43,45){\LARGE{$\textcolor{white}{\blacksquare}$}}
\put(47,46){$\Gamma$}
\end{overpic}
}
\hspace{10pt}
\subfigure[\ $k_x - k_z$ plane]{
\begin{overpic}[height=0.4\columnwidth]
{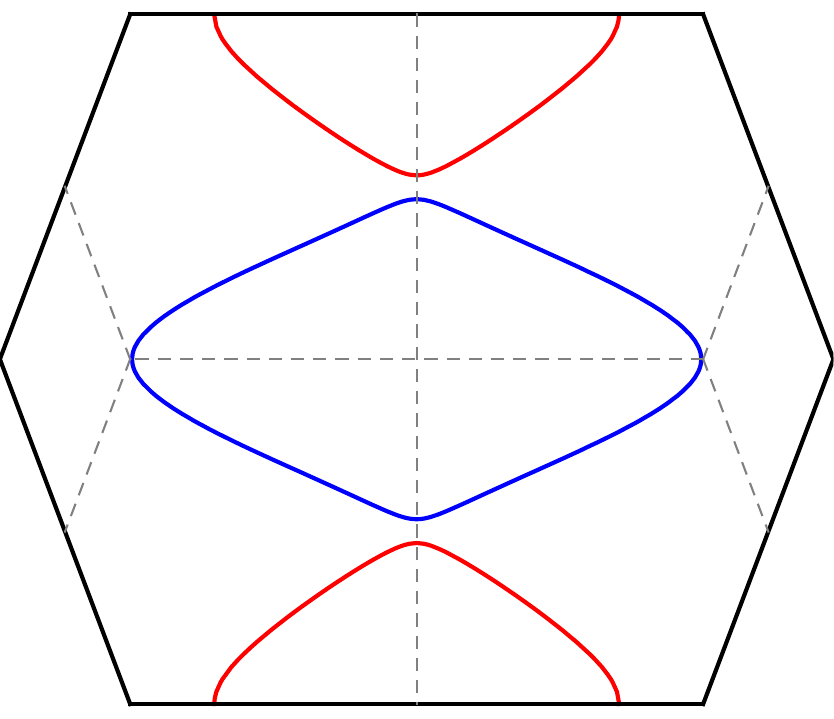}
\put(52,44){\LARGE{$\textcolor{white}{\blacksquare}$}}
\put(55,46){$\Gamma$}
\put(55,100){$Z$}
\end{overpic} 
}
\caption{\label{fermi-surfaces} (Color online) 
(a) View of the Fermi surfaces in the $k_x$-$k_y$ plane. The hole-like Fermi surface (red) is centered
about $k_z = 2\pi/c$ while the electron-like Fermi surface (blue) is centered about $\Gamma$.
(b) View of the Fermi surfaces in the $k_x$-$k_z$ plane.}
\end{figure}
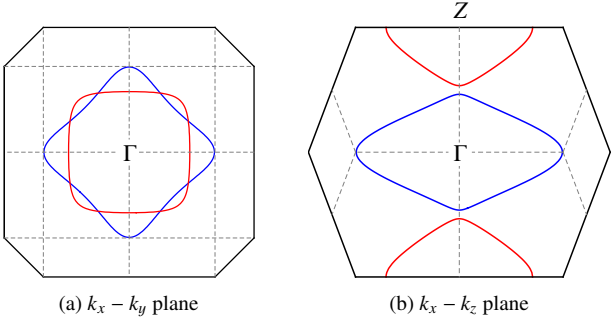

\section{Model}
The heavy-fermion physics of URu$_2$Si$_2$ 
is signalled by a peak in the magnetic susceptibility\cite{mydosh-1985}
indicating a coherence temperature $\sim 60K$, well above the
hidden order transition temperature $T_O$. It is reasonable
to assume that near the HO phase the relevant physics is encompassed 
by that of a (heavy) Fermi liquid. 
Recent work\cite{oppeneer-dft,oppeneer-dft-2,fujimoto-nematic,softarpes} supports this,
suggesting itinerant character for the $\U$ $5f$ electrons, with coherent $\U$
dominant quasi-particles near the Fermi surface, can account for 
many of the properties of the paramagnetic state.

\begin{figure}[t]
\subfigure[\ $t_{12}$,$t_a$]{
  \includegraphics[width=0.3\columnwidth]{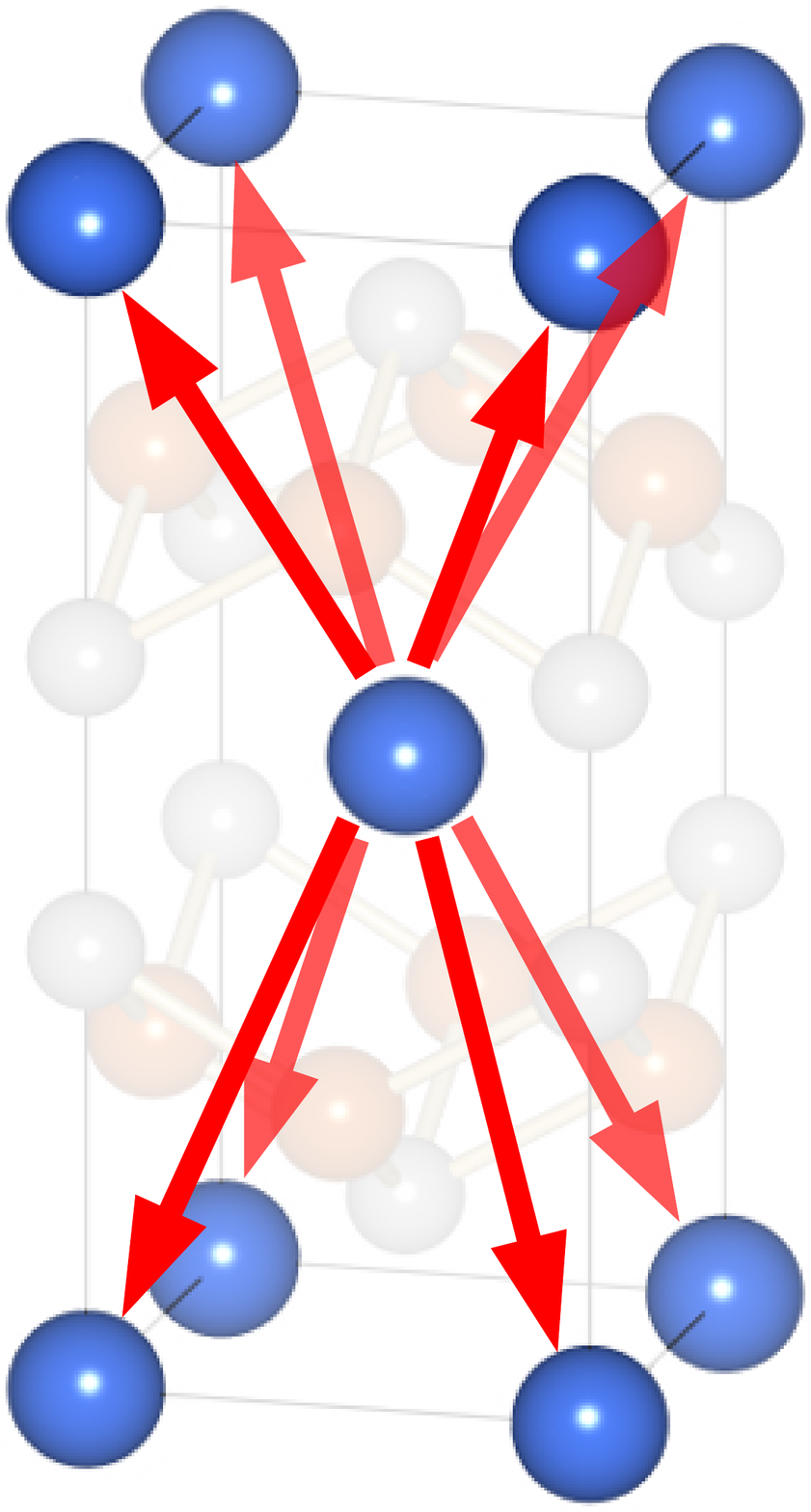}
}
\subfigure[\ $t'_a$]{
  \includegraphics[width=0.3\columnwidth]{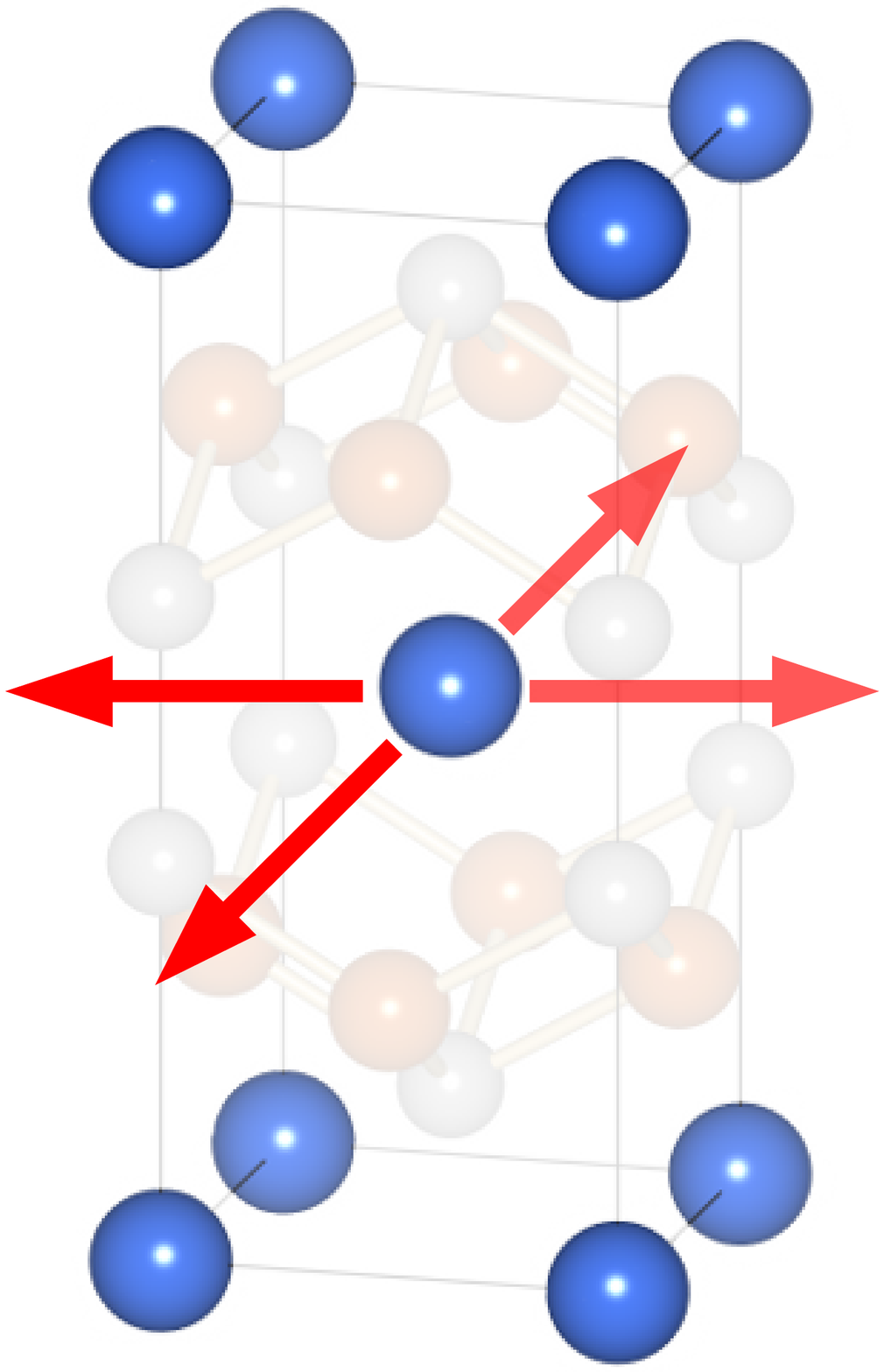}
}
\subfigure[\ $t''_a$]{
  \includegraphics[width=0.3\columnwidth]{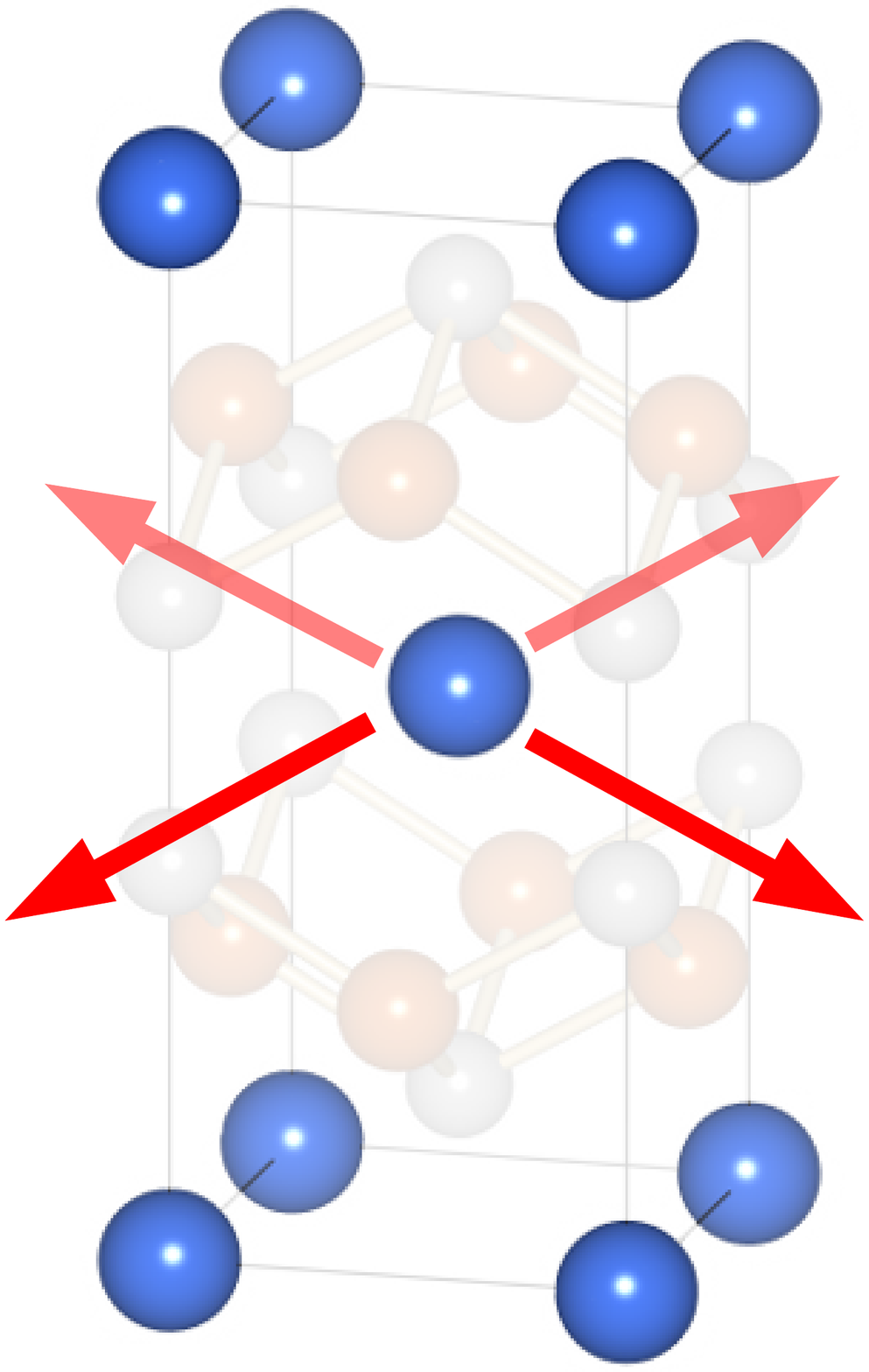}
}
\caption{\label{hoppings}
(Color online) 
Graphical represenation of the relevant hopping amplitudes along the (a)
body diagonal (b) in-plane axes and (c) in-plane diagonals shown in the
unit cell of URu$_2$Si$_2$.
}
\end{figure}

Motivated by this, we consider single particle $5f$ states in
a crystal field of $D_{4h}$ symmetry, using the notation of Ref. \cite{thalmeier-landau} for
the irreducible representations of $D_{4h}$.
The strong atomic spin-orbit coupling splits the $5f$ levels into a low-lying $j=5/2$ sextet and
 high-lying $j=7/2$ octet. Since the $j=7/2$ states lie significantly higher in energy than
the $j=5/2$, we keep only the latter in our discussion. Under the potential
of the crystal lattice this sextet spilts into
three Kramers doublets, as $\Gamma_{5/2} = 2\Gamma_7 \oplus \Gamma_6$, explicitly
\begin{eqnarray*}
  f_{1\pm} &=& \cos{\theta}\ f_{\pm \frac{5}{2}} + \sin{\theta}\ f_{\mp \frac{3}{2}} \\
  f_{2\pm} &=& -\sin{\theta}\ f_{\pm \frac{5}{2}} + \cos{\theta}\ f_{\mp \frac{3}{2}} \\
  f_{3\pm} &=& f_{\pm\frac{1}{2}}
\end{eqnarray*}
We consider a tight-binding model for URu$_2$Si$_2$ using states with $\Gamma^{(1)}_7$, $\Gamma^{(2)}_7$ and
$\Gamma_6$ character, including contributions from
nearest and next-nearest neighbor hoppings. 
The $\Gamma_6$ bands are
involved in producing the incommensurate inelastic neutron                             
peak\cite{broholm-1991},                                                             
at a nesting wave-vector connecting $\Gamma_7$ and $\Gamma_6$, through the
gapping of $\Gamma_7$ Fermi surface in the ordered phases.
For the sake of simplicity we focus only on Fermi                                   
surfaces of $\Gamma_7$ character as they form the superspin of the current              
study. If we include hoppings along the body diagonals, as 
well as in along the in-plane axes and diagonals symmetry
restrict us to a Hamiltonian of the form
\begin{eqnarray}
  \label{tight-binding-model}
  H_0 &=& \sum_{k\sigma}\left( 
                          A_{1k} \h{f}_{1\sigma,k}f_{1\sigma,k} + 
                          A_{2k} \h{f}_{2\sigma,k}f_{2\sigma,k}
                        \right)+ \nonumber \\    
    & & \sum_k\left(
                    C_k\h{f}_{1+,k}f_{2+,k} + 
                    \cc{C}_k\h{f}_{1-,k}f_{2-,k} + 
                    {\rm h.c}
                 \right)+ \nonumber \\
    & & \sum_k\left(
                    D_k\h{f}_{1+,k}f_{2-,k} - 
                    \cc{D}_k\h{f}_{1-,k}f_{2+,k} + 
                    {\rm h.c}
                 \right),
\end{eqnarray}
where $\h{f}_{\alpha\sigma}$, with $\alpha=1,2$, $\sigma=\pm$ creates a state in the $\Gamma^{(\alpha)}_7$ doublet
with pseudo-spin $\sigma$. To obtain a Fermi surface that matches our
criteria above, we find it possible to take $C_k=0$. The remaining
dispersion functions take the form
\begin{eqnarray*}
  A_{\alpha k} &=&
  8 t_a \cos{\left(\frac{ak_x}{2}\right)}\cos{\left(\frac{ak_y}{2}\right)}\cos{\left(\frac{ak_z}{2}\right)} +\\
&&2t'_a \left(\cos{\left(ak_x\right)} + \cos{\left(ak_y\right)}\right) + \\
&&4t''_a \cos{\left(a k_x\right)}\cos{\left(a k_y\right)} - \mu +{\rm sgn}(\alpha)\frac{\Delta_{12}}{2},\\
D_k &=& 4 t_{12}\left[\sin{\left(\frac{a(k_x+k_y)}{2}\right)} - i \sin{\left(\frac{a(k_x-k_y)}{2}\right)}\right]\sin{\left(\frac{c k_z}{2}\right)},
\end{eqnarray*}
where $\Delta_{12}$ is the effective crystal 
field splitting between $\Gamma^{(1)}_7$ and $\Gamma^{(2)}_7$ levels, ${\rm sgn}(\alpha) =\pm 1$ for $\alpha=1,2$
and $c$ and $a$ are the lattice constants. Physically, $t_a$ represents hopping
along the $a(\hat{x}+\hat{y}) + \frac{c}{2}\hat{z}$ and equivalent directions, $t'_a$  along $a\hat{x}$ and $a\hat{y}$
and $t''_a$ along $a(\hat{x}+\hat{y})$ and equivalent directions, as shown in Fig. \ref{hoppings}.
In the following we fix
hoppings $t_1 = t_2 = -0.3$, $t'_1=-0.87$, $t'_2=0.0$, $t''_1=0.375$, $t''_2 = 0.25$ with
chemical potential $\mu =-0.5$ and crystal field splitting $\Delta_{12} = 3.5$. For concreteness
we take $|t_{12}|=0.7$ in the Fermi surface plots, but will allow it vary when discussing phenomenological
models. We note that Fermi surface does not depend strongly on the choice
of $t_{12}$.
Geometrically, one expects this $t_{12}$ hopping to be related
to hybridization with the Ru $d$ orbtials.
As we have neglected several bands from our description, 
we treat the remainder of the system as
a charge resevoir, keeping the chemical potential
fixed and allowing the density vary. We note that for
$t_{12}=0$ this tight-binding model has global ${\rm SU}(2)$ pseudo-spin
rotation symmetry within each doublet. For finite $t_{12}$ this is broken to a single extra ${\rm U}(1)$ symmetry between the $\Gamma^{(1)}_7$
and $\Gamma^{(2)}_7$ doublets, transforming $f_{1\sigma} \rightarrow e^{i\sigma\psi}f_{1\sigma}$ and
$f_{2\sigma}\rightarrow e^{-i\sigma\psi} f_{2\sigma}$.

Using the above hopping parameters, the Fermi surface
shown in Fig. \ref{fermi-surfaces} is obtained. These 
parameters have been chosen to respect the following experimental
and theoretical results. From 
Hall and magnetoresistivity measurements, one expects closed electron and hole
Fermi surfaces of nearly equal size.
Ab-initio calculations
suggest that these electron and hole pockets are composed mainly of
$\Gamma^{(1)}_7$ and $\Gamma^{(2)}_7$. Furthermore, 
the similarity of the Fermi surfaces in the HO and AF states
and nesting between the electron and hole Fermi surfaces at
the AF ordering vector $\vecc{Q}$, strongly suggests that the HO and AF orderings
occur at the same wave-vector. Under this assumption we then need the
nesting to be imperfect to match the small pockets observed in oscillation
measurements\cite{amitsuka-1999,nakashima-2003}. This is implemented
in a fashion consistent with ab-initio results, with the electron pocket
near $\Gamma$ being elongated in the $[100]$ and $[010]$ directions and
the hole pocket at $Z$ being elongated in the $[110]$ and $[1\cb{1}0]$ 
directions. After folding along $\vec{Q}$ one then has four small pockets
along $[100]$ and $[010]$ as shown in Fig \ref{fermi-surface-gapping-xy}.

\begin{figure}[b]
\subfigure{
  \includegraphics[width=0.47\columnwidth]{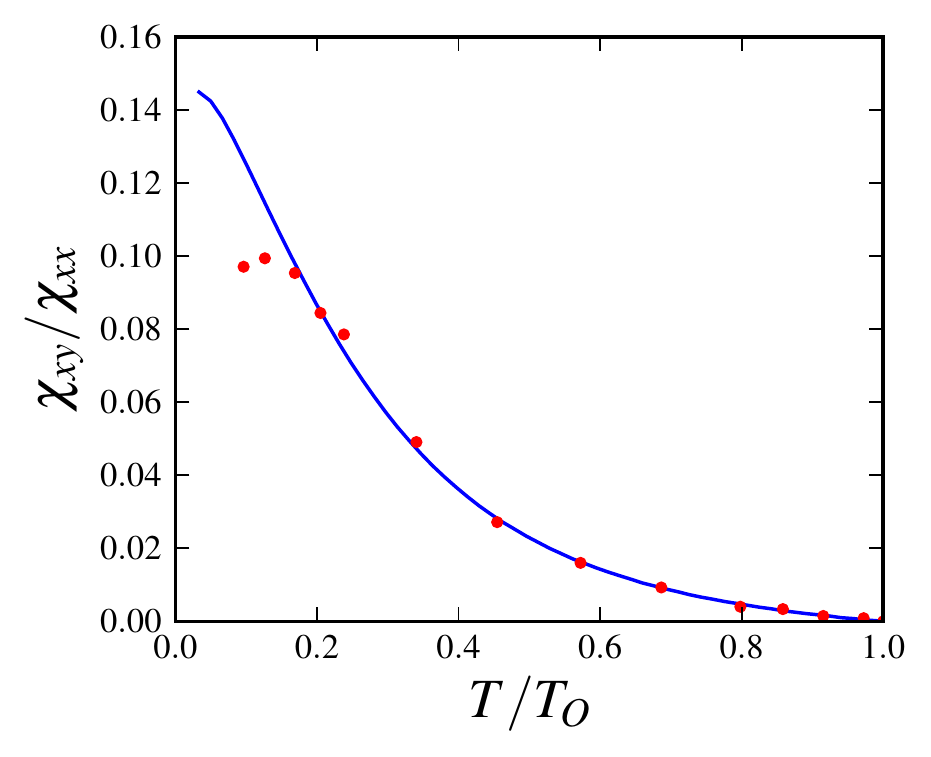}
}
\subfigure{
  \includegraphics[width=0.47\columnwidth]{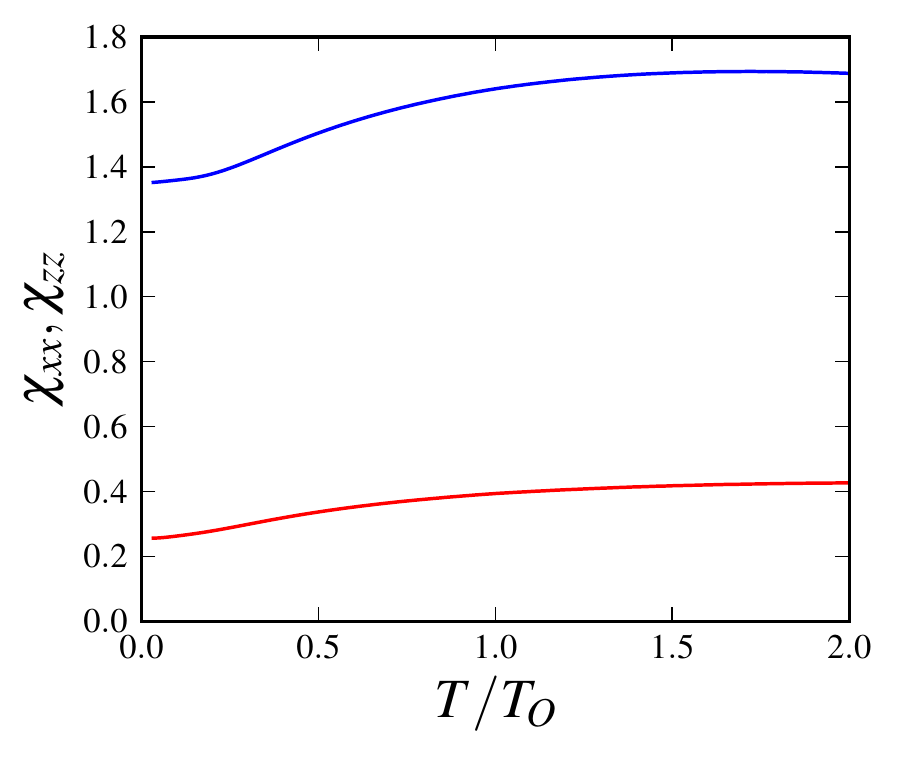}}

\caption{
\label{susceptibility} 
(Color online) 
(a) The off-diagonal magnetic susceptibility
$\chi_{xy}$ in the $E(1,1)$ at a function of temperature. One can obtain a strong resemblance
to the
measured susceptibility in \cite{okazaki-torque}, shown in red, by fitting the overall
amplitude of $\chi_{xy}/\chi_{xx}$ and the position of $T_O$.
(b) The uniform magnetic susceptibility in the $E(1,1)$ phase
 as a function of temperature. Both $\chi_{xx} = \chi_{yy}$ (red) and $\chi_{zz}$ (blue) are shown.}
\end{figure}
\section{Order parameter} 
The key signals found in the torque measurements are $\chi_{xy} \neq 0$
and $\chi_{xx}=\chi_{yy}$ within the HO phase.
Landau-Ginzburg arguments\cite{thalmeier-landau} show
that an order parameter consistent with these facts and their temperature dependence
 should transform as the two-dimensional $E$ irreducible
representation\footnote{This representation transforms
as $\sim (xz,yz)$} of the $D_{4h}$ point group, with periodicity at the wave vector $\vecc{Q}$.
The details of the torque oscillations single out a set
of symmetry related directions
in this two dimensional space, namely the four orientations $(\pm 1,\pm 1)$, 
so we will denote this type of order parameter as $E(1,1)$.
In the picture outlined above,
where the relevant degrees of freedom are the two $\Gamma_7$ bands,
there are only two local $E$ type order parameters that mix $f_1$
and $f_2$, taking advantage of the nesting at $\vecc{Q}$. These are 
the $x$ and $y$ components of the vectors $i\left(\vecc{S}_{12} -\vecc{S}_{21}\right)$
or $\vecc{S}_{12} + \vecc{S}_{21}$.

The first $E$ type order parameter is
inconsistent with resonant X-ray scattering
results \cite{floquet-2011} as it carries an electric quadrupole
moment. The second breaks time-reversal and, as the in-plane components
of the magnetic moment operator also transform as $E$, 
an induced magnetic moment is expected generically.
This can be seen explicitly by
restricting to the $\Gamma_7$ doublets and constructing the moment
operator $\vecc{\mu} =\vecc{L}+2\vecc{S}$ at each site in the $\Gamma_7$ basis
\begin{eqnarray} \nonumber
  \mu_x &=& \frac{6\sqrt{5}}{7}
  \left[
    \sin{(2\theta)} \left(S^x_{11}-S^x_{22}\right)+
    \cos{(2\theta)} \left(S^x_{12} + S^x_{21}\right)
    \right], \\ 
\nonumber
  \mu_y &=& \frac{6\sqrt{5}}{7}
  \left[
    \sin{(2\theta)} \left(S^y_{11}-S^y_{22}\right)+
    \cos{(2\theta)} \left(S^y_{12} + S^y_{21}\right)
  \right], \\\nonumber
\mu_z &=& 
\frac{24}{7}\left[
\cos{(2\theta)}\left(S^z_{11}-S^z_{22}\right)-
\sin{(2\theta)}\left(S^z_{12}+S^z_{21}\right)\right] \\ 
&& +\frac{6}{7}\left[
S^z_{11}+S^z_{22}
\right].
\label{magnetic-moment}
\end{eqnarray}
 For $\theta = \frac{\pi}{4} + \delta$ with $\delta$ small,
the terms proportional to $\vecc{S}_{12} + \vecc{S}_{21}$ 
in the $x$ and $y$ components are $O(\delta)$, while
the $z$ component remains $O(1)$.
One has the scenario, that for small $\delta$ 
the magnetic coupling to this type of $E$
order parameter is significantly reduced\footnote{
Note that this mechanism does not work for the other $E$ type
order parameter,
as there is no other $E$ type order parameter with the
required quadrupole moments within the $\Gamma_7$ doublets.
}. 
For example the neutron scattering cross-section
is $O(\delta^2)$. For these reasons
we take the $x$ and $y$ components of Eq. \ref{order-parameter} as the HO
with the $z$ component as the AF \footnote{A similar superspin 
relation between the HO and AF was suggested in Ref. \cite{kotliar-2011}
where the HO is hexadecapolar and related to
the AF through a $U(1)$ transformation. This hexadecapolar order
parameter does not break tetragonal
symmetry and is thus inconsistent with the torque experiment.}.
Note that even though the expectation is supressed,
the presence of other terms $\sim \sin{(2\theta)}$ in the moment operator allows terms
of $O(1)$ to present in the in-plane susceptibilities.
The bare susceptibilities in the $E(1,1)$ phase are shown in Fig. \ref{susceptibility}.
The temperature dependence of the the $\chi_{xy}$ signal is strikingly similar
to that found in the torque experiments. Deviations at the lowest temperatures
are likely due to effects from the superconducting phase.

\section{Phenomenological theory}
Due to the approximate nesting of the Fermi surfaces,
the similarity to the AF state and
the implications of the Landau-Ginzburg analysis of the torque experiments\cite{thalmeier-landau},
we consider only ordering at wave-vector $\vecc{Q}$.
To explore this scenario, 
consider the phenomenological model, where the tight-binding 
model is supplemented by the coupling terms
\begin{eqnarray}
  H_\phi &=& -\frac{\vecc{\phi}}{\sqrt{2}} \cdot \sum_i e^{i\vecc{Q}\cdot\vecc{r}_i} \left(
\vecc{S}_{12}(i) + \vecc{S}_{21}(i)
\right), \\ 
&=& -\frac{\vecc{\phi}}{2\sqrt{2}} \cdot \sum_k \left(
\h{f}_{1,k} \vecc{\sigma} f_{2,k+Q}+
\h{f}_{2,k} \vecc{\sigma} f_{1,k+Q}
+{\rm h.c.}
\right), \nonumber
\end{eqnarray}
where the momentum sum runs over the reduced Brillouin zone,
folded along $\vecc{Q}$, as well as the quadratic terms
\begin{equation}
  \label{mf-quad}
    N\left(\frac{1}{2 g_{xy}} \left(\phi_x^2+\phi_y^2\right) + \frac{1}{2 g_z} \phi_z^2\right),
\end{equation}
where we have introduced the couplings $g_{xy}$ and $g_{z}$.
 The $E(1,1)$ phase cooresponds to
$|\phi_x| = |\phi_y| \equiv \phi$ and $\phi_z=0$. The phase where $\phi_x = \phi_y=0$
and $\phi_z \neq 0$ is magnetic, without any suppression
factors of $\delta$ in the magnetic coupling, and does not break the
$C_4$ symmetry present in the point group. We identify
this with the pressure induced AF phase

\begin{figure}[b]
\subfigure[\ $\phi=0.75$]{
\begin{overpic}[width=0.3\columnwidth]
{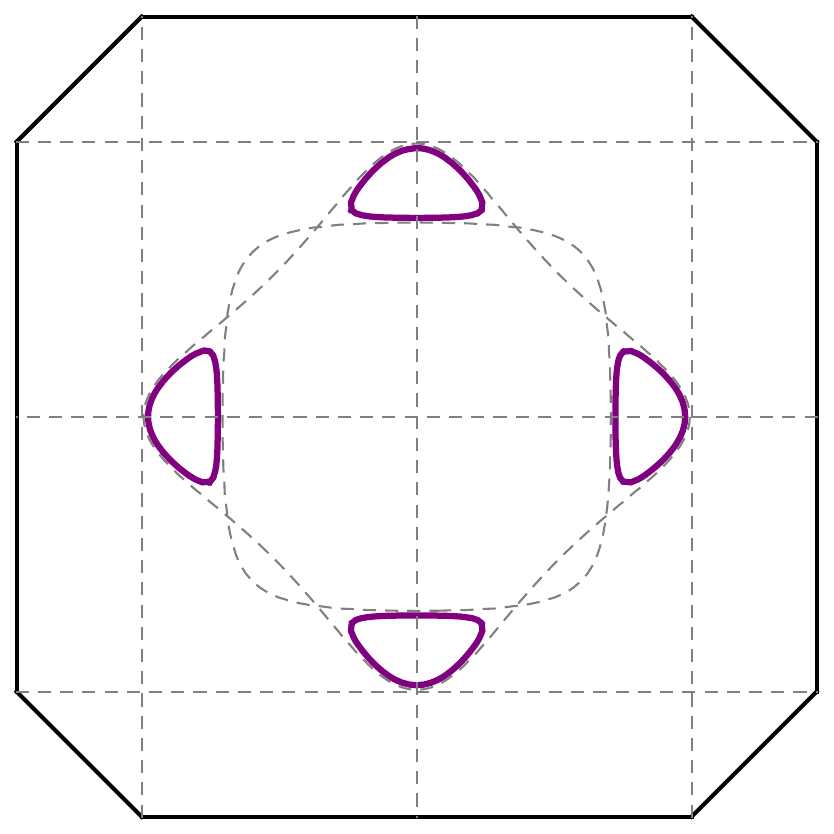}
\end{overpic}
}
\subfigure[\ $\phi=0.53$]{
\begin{overpic}[width=0.3\columnwidth]
{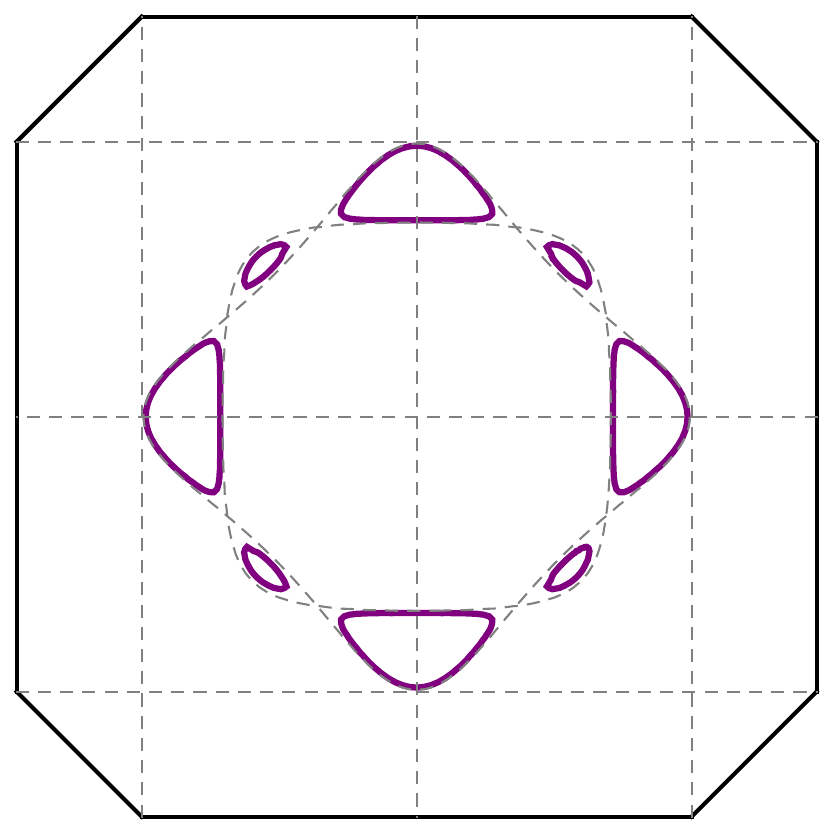}
\end{overpic}
}
\subfigure [\ $\phi=0.167$]{
\begin{overpic}[width=0.3\columnwidth]
{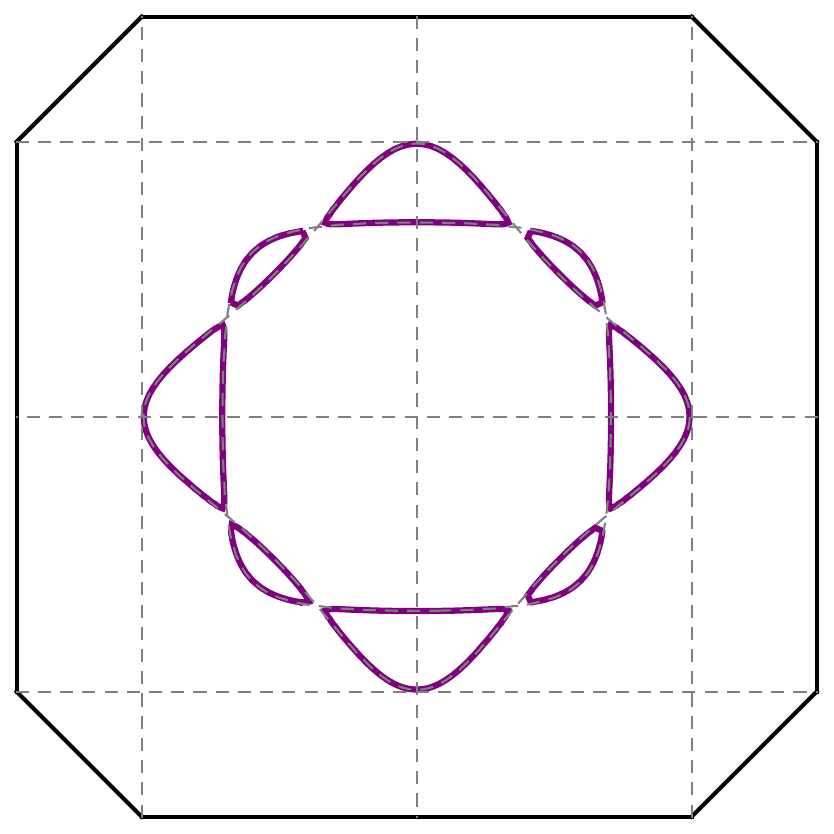}
\end{overpic}
}
\caption{
\label{fermi-surface-gapping-xy}
(Color online) 
Fermi surface after imposition of finite $\phi = |\vecc{\phi}|$ in the
$k_z = 0$ plane for several values of $\phi$ (in the unfolded Brillouin zone).
}
\end{figure}

Explicitly, we first write the kinetic part
of the Hamiltonian as
\begin{eqnarray}
H_0 &=&  \sum_k 
  \h{\Psi}_k \left(
    \begin{tabular}{cccc}
      $A_{1k}$ & $0$ & $0$ & $D_k$ \\ 
      $0$ & $A_{1k}$ & $-\cc{D}_k$ & $0$ \\ 
      $0$ & $-D_k$ & $A_{2k}$ & $0$ \\ 
      $\cc{D}_k$ & $0$ & $0$ & $A_{2k}$ \\ 
      \end{tabular}
      \right) \Psi_k \\
      &\equiv&
\sum_k 
  \h{\Psi}_k \left(
    \begin{tabular}{cc}
      $A_{1k}$ & $\gamma_k$ \\
      $\h{\gamma}_k$ & $A_{2k}$\\
      \end{tabular}
      \right) \Psi_k
\end{eqnarray}
where $\h{\Psi}_k = ( \h{f}_{1+,k}\ \h{f}_{1-,k}\ \h{f}_{2+,k}\ \h{f}_{2-,k})$ and $\gamma_k = i\sigma_y {\rm Re}D_k+i\sigma_x{\rm Im}D_k$.  
Then the phenomenological Hamiltonian
is given by
\begin{equation}
\label{full-ham}
H =  \sum_{k}'
    \h{\left(
    \begin{tabular}{c}
      $\Psi_k$ \\
      $\Psi_{k+Q}$ \\
      \end{tabular}\right)}
 \left(
    \begin{tabular}{cccc}
      $A_{1k}$ & $\gamma_k$ & $0$ & $-\frac{\vec{\phi} \cdot\vec{\sigma}}{2\sqrt{2}}$ \\
      $\h{\gamma}_k$ & $A_{2k}$ & $-\frac{\vec{\phi} \cdot\vec{\sigma}}{2\sqrt{2}}$ &$0$\\
      $0$  & $-\frac{\vec{\phi} \cdot\vec{\sigma}}{2\sqrt{2}}$ & $A_{1k+Q}$ & $\gamma_{k+Q}$\\
       $-\frac{\vec{\phi} \cdot\vec{\sigma}}{2\sqrt{2}}$ & $0$ & $\h{\gamma}_{k+Q}$ & $A_{2k+Q}$
    \end{tabular}
    \right) 
    \left(
    \begin{tabular}{c}
      $\Psi_k$ \\
      $\Psi_{k+Q}$ \\
      \end{tabular}\right)
\end{equation}
where the momentum sum now runs over the reduced Brillouin zone. 
In
this general form one cannot access the spectrum analytically of Eq. \ref{full-ham}, but
 since for $k_z=0$ the
off-diagonal term $D_k$ is zero, one can find the spectrum in this plane, 
yielding branches $E^{\pm}_k$ and $E^{\pm}_{k+Q}$ where
\begin{equation*}
  E^{\pm}_k = \frac{A_{1k} + A_{2,k+Q}}{2} \pm\sqrt{
\left(\frac{A_{1k} - A_{2,k+Q}}{2}\right)^2 + \left|\frac{\vecc{\phi}}{2\sqrt{2}}\right|^2}
\end{equation*}
This implies that the Fermi surfaces have the same cross
sections at $k_z=0$, independent of the orientation of $\vecc{\phi}$. 
This  naturally explains the similarity in the oscillation
measurements\cite{amitsuka-1999,nakashima-2003} in the HO and AF phases
and the remaining small pockets. The Fermi
surface for various values of $|\vecc{\phi}|$
is shown in Fig. \ref{fermi-surface-gapping-xy}.
As the ordering is strengthened, the remaining Fermi surfaces form four
small pockets along the $x$ and $y$ axes. 
This removal of large sections of the Fermi surface within the HO
phase is qualitatively consistent with implications from transport and specific heat measurements \cite{mydosh-1985}.
The destruction of the pockets along
the diagonal directions occurs due to more robust nesting in planes
along the intersection points of the electron and hole Fermi surfaces.
A similar feature has been noted in recent ab-initio Fermi
surfaces \cite{oppeneer-dft-2}.

\begin{figure}[t]
\subfigure{
  \includegraphics[width=0.47\columnwidth]{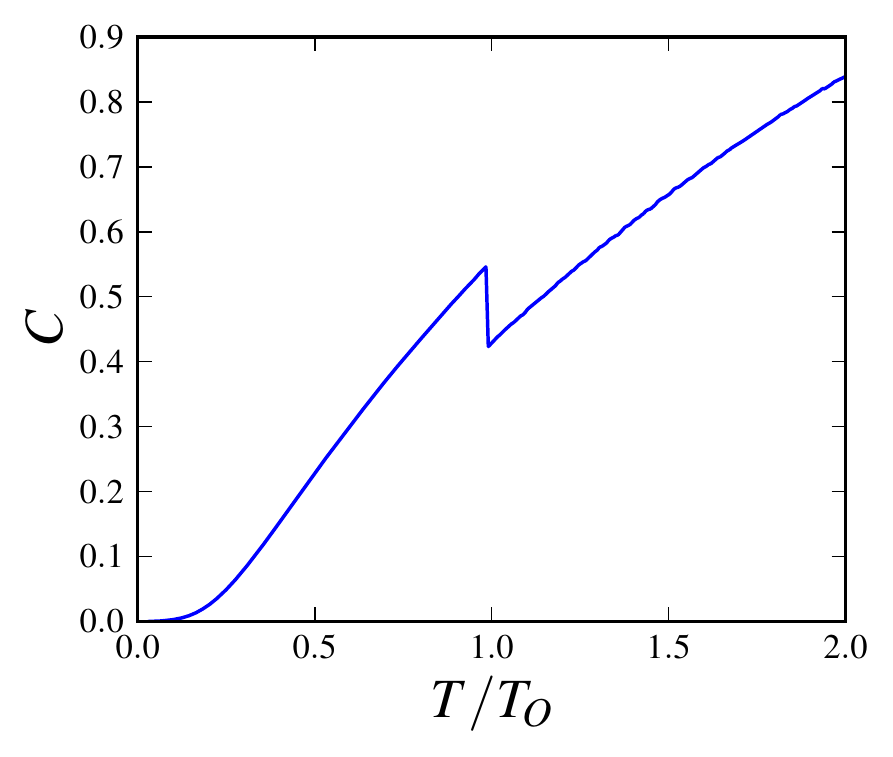}
}
\subfigure{
  \includegraphics[width=0.47\columnwidth]{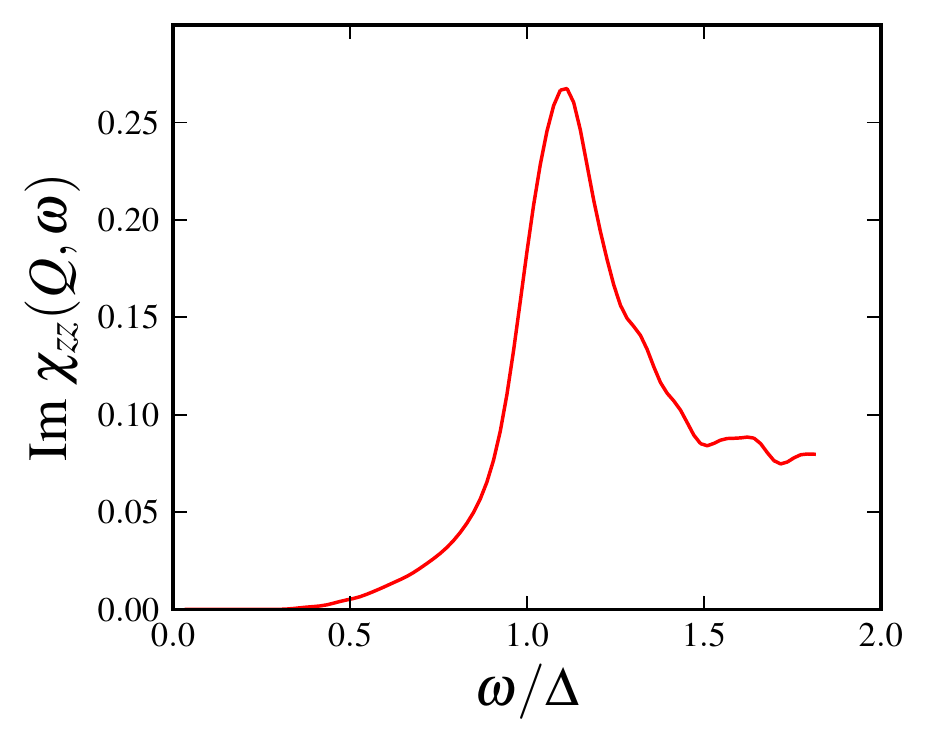}
}
\caption{ 
\label{im-susceptibility}
(Color online) 
(a) The specific heat as a function of temperature. Note the jump at $T_O$.
(b) The imaginary
part of the susceptibility in the $zz$ channel at $\vecc{Q}$ at $T=0$. 
The frequency $\omega$ is scaled by the gap at perfect nesting 
$\Delta = |\vecc{\phi}|/\sqrt{2}$.}
\end{figure}

A key signature of the
HO phase is the inelastic neutron scattering peak at $\vecc{Q}$.
To examine this the
imaginary part of the bare susceptibility at the ordering vector $\vecc{Q}$, $\im \chi_{zz}(\vecc{Q},\omega)$
is computed 
and shown in Fig. \ref{im-susceptibility} (a). There
is a clear gap below a peak at finite frequency, reminiscent of the neutron results. 
The specific heat shown in Fig. \ref{im-susceptibility} (b) also shows a clear jump              
at $T_O$ as expected, but the calculated value is too small to be realistically compared with the 
experimental results.                                                                 
The precise size of jump in the model depends on the strength of order parameter, $\phi$,
tendency of nesting on the Fermi surface, interaction effects and details of the some the other
bands that have been left out of our model (for example from the gapping of the
incommensurate modes).

\begin{figure}[t]
\subfigure{
  \includegraphics[width=0.47\columnwidth]{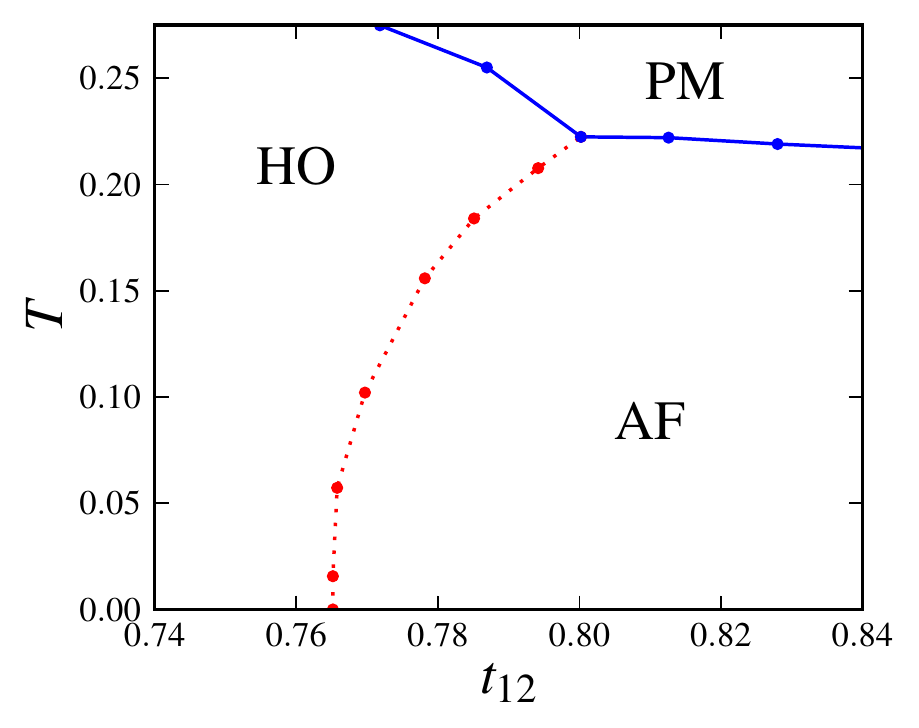}
}
\subfigure{
  \includegraphics[width=0.47\columnwidth]{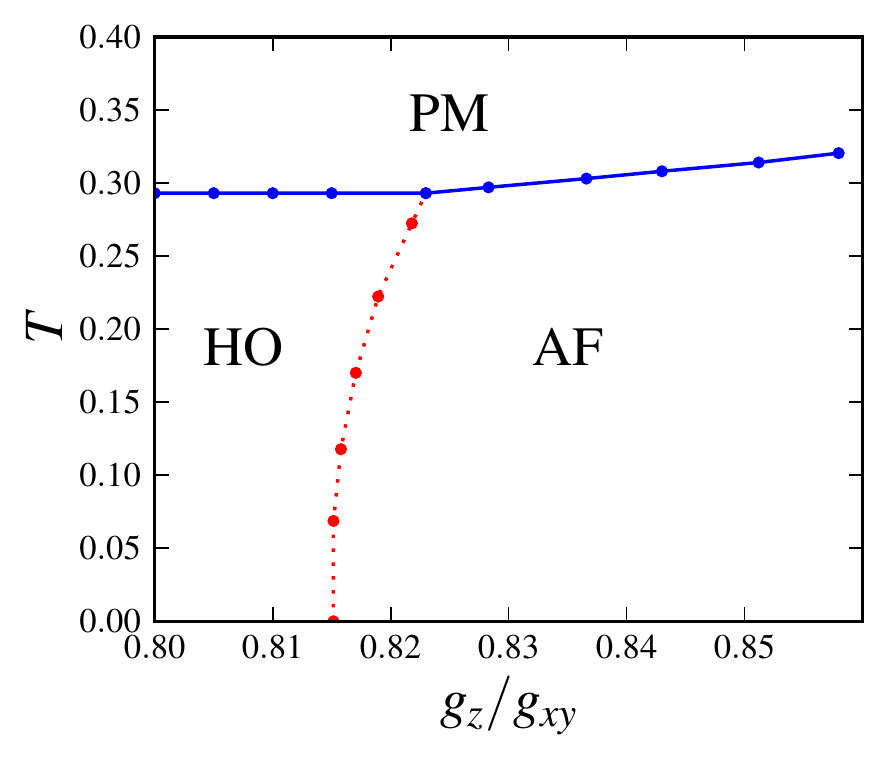}
}
\caption{\label{phase-diagram}
(Color online) 
(a) Mean field phase diagram as a function of the hopping parameter
$t_{12}$, which breaks the HO-AF degeneracy, at $g_z / g_{xy} = 0.8$.
(b) Phase diagram as a function of the coupling constant ratio $g_z / g_{xy}$
where $t_{12}=0.7$ and $g_{xy}=5.0$.
}
\end{figure}

To study the transition for the HO to AF phases we look at the mean field
phase diagram as a function of temperature, effective coupling constants $g_{xy}$ and $g_z$ and $t_{12}$, the 
hopping which breaks pseudo-spin rotation symmetry. 
One can evaluate the free energy for this Hamiltonian numerically,
including the quadratic parts of Eq. \ref{mf-quad}, and minimize to
find the value of $\vec{\phi}$.
As the locking of the pseudo-spin
orientation to the spatial orientation is controlled by the $t_{12}$ hopping,
to obtain an $E(1,1)$ phase over a $E(1,0)$ or $E(0,1)$ phase we choose the phase
of $t_{12}$ to be $\pi/4$. This is supported by Slater-Koster type calculations
of $t_{12}$, which also yield a phase of $\pi/4$.
We find that for any finite value of $t_{12}$ with equal couplings,
$g_{xy}=g_z$, the AF state is favoured. This allows for a scenario where $g_{xy}>g_z$ and $t_{12}$ increases
as pressure is raised from ambient, causing a transition from HO to AF as shown in Fig. \ref{phase-diagram} (a).
A more direct possibility is allowing $g_z/g_{xy}$ to vary from $g_{xy}>g_z$ at
ambient pressure to $g_z > g_{xy}$ at higher pressure, and fixing $t_{12}$, as shown in Fig. \ref{phase-diagram} (b).
In both these cases the HO to AF transition is first order, and the qualitative topology is similar
to that seen as a function of pressure and temperature in URu$_2$Si$_2$.

\section{Discussion} 
An important ingredient in this proposal is the
fine-tuning of the crystal field angle $\theta$ to 
be near enough to $\pi/4$ to have the basal plane moments avoid detection thus far. 
Depending on the magnitude of the
deviation $\delta$, the order parameter should be observable in
polarized neutron scattering. This leads to a prediction 
of small moment in the basal plane in the HO phase,
oriented along $[110]$.  So far, neutron scattering has been focused on        
scattering at wavevector $(1,0,0)$, favourable to
detecting ordering oriented along the $z$ axis.
Furthermore, when (un-polarized) scattering is done at wave-vector $(1,0,0)$ 
the small moment from the HO would be lost in the signal from the parasitic
AF moment.
To avoid a signal from the parasitic moment present
in the HO phase, while still detecting a small
moment along $[110]$,
we suggest a careful neutron scattering analysis with wave-vector $(0,0,1)$ (parallel
to $\vecc{Q}$) since the
$z$ component moment would not contribute                                             
due to the geometric factor in the            
scattering cross section.                                                           
In particular, when neutron is polarized
parallel to $\vecc{Q}$, all magnetic scattering is spin-flip and
should come from the basal plane moment. Polarized scattering
along $(1,0,0)$ should also be effective at uncovering the moment,
so long as the sensitivity is sufficiently high.
While the quantitative determination of the angle $\theta$, and thus
expected moment size to be seen in experiment, is beyond the scope
of the current study recent LDA+U calculations \cite{ikeda-unpublished}
suggest that such fine-tuning could be realized in URu$_2$Si$_2$.
Effects of high magnetic fields could also lead to testable
implications of this model. One expects that at large enough fields
this will destroy the HO due to breaking of
nesting between the $\Gamma^{(1)}_7$ and $\Gamma^{(2)}_7$ orbitals.
 The complex structure of the magnetic moment operator
in Eq. \ref{magnetic-moment} could lead non-trivial Fermi surface
splittings, perhaps leading way to another type
of ordering. The relation of this to the phases\cite{mydosh-2007} which appear in the
vicinity of $\sim 37{\rm T}$ warrants a detailed study, 
which we leave to future work.

In summary, we have presented
a simple effective tight-binding model for the relevant bands in URu$_2$Si$_2$.
Motivated from this model, we propose that the HO is realized as a rank-5
inter-orbital spin-density wave oriented along the $[110]$ direction. This
gives the correct breaking of tetragonal symmetry to account for the
recent torque results\cite{okazaki-torque}.
A natural relation between the HO and AF phases is provided, connecting them
through a pseudo-spin rotation as well as explaning the similarity of their
Fermi surfaces. The Goldstone mode in the HO is gapped due
to terms that break pseudo-spin rotational symmetry, manifesting as
the finite frequency peak seen in inelastic neutron scattering experiments.
The pressure-temperature phase diagram is also understood through this symmetry
breaking.
While the size of the moment in
the basal plane in the HO can be suppressed
by fine-tuning of the crystal field strengths,
it should be detectable by polarized neutron scattering.
 
\emph{Note added:} After completion of this work, we have been
made aware of a first principles LDA+U study \cite{ikeda-unpublished} which
has proposed an time-reversal breaking $E$-type order parameter 
as a strong candidate for the HO phase in URu$_2$Si$_2$. The smallness of 
the moment in this work lends support to our current proposal.

\acknowledgements We thank Christoph M. Puetter and
Jeff Asaf Dror for useful discussions. This work was supported by the NSERC of Canada.
\bibliography{hidden-order}

\end{document}